# GJ3470-d and GJ3470-e: Discovery of Co-Orbiting Exoplanets in a Horseshoe Exchange Orbit

Phillip Scott[1], Jaxon Taylor[2], Larry Beatty[3], Jim Edlin[4], Phil Kuebler[4], Mike Dennis[4], David Higgins[4], Alberto Caballero[5], Alberto Garcia[6]

21 April, 2023

[1]OKSky Observatory, OK, USA
[2]Jaxon Taylor Observatory, OK, USA
[3]New Mexico Skies Observatory, NM, USA
[4]Cheddar Ranch Astronomical Group, OK, USA
[5]Habitable Exoplanet Hunting Project, Spain
[6]Rio Cofio Observatory, Spain

Abstract

We report the discovery of a pair of exoplanets co-orbiting the red dwarf star GJ3470. The larger planet, GJ3470-d, was observed in a 14.9617-days orbit and the smaller planet, GJ3470-e, in a 14.9467-days orbit. GJ3470-d is sub-Jupiter size with a 1.4% depth and a duration of 3 hours, 4 minutes. The smaller planet, GJ3470-e, currently leads the larger planet by approximately 1.146-days and is extending that lead by about 7.5-minutes (JD 0.0052) per orbital cycle. It has an average depth of 0.5% and an average duration of 3 hours, 2 minutes. The larger planet, GJ3470-d, has been observed on seven separate occasions over a 3-year period, allowing for a very precise orbital period calculation. The last transit was observed by three separate observatories in Oklahoma and Arizona. The smaller planet, GJ3470-e, has been observed on five occasions over 2-years. Our data appears consistent with two exoplanets in a Horseshoe Exchange orbit. When confirmed, these will be the second and third exoplanets discovered and characterized by amateur astronomers without professional data or assistance. It will also be the first ever discovery of co-orbiting exoplanets in a Horseshoe Exchange orbit.

# 1 Introduction

GJ3470 is an M type Red Dwarf star located in the constellation of Cancer. It already has one confirmed exoplanet, GJ3470-b, which is a Neptune-like exoplanet 13.9 times the mass of the Earth, 0.408 times the radius of Jupiter, and an orbital period of 3.3 days (NASA, 2020). There is also one candidate planet, GJ3470-c, that was published on Arxiv.org (Scott, 2020).

Alberto Caballero created The Habitable Exoplanet Hunting Project (Caballero, 2010), which included more than 30 amateur observers from 10 countries around the world. The project was intended to observe selected nearby red dwarf stars within 100 light years that included at least one confirmed exoplanet. The hope was that by concentrating observations 24/7 for a specific period of time, that potentially habitable planets may be found. This resulted in the discovery of GJ3470-c, a Saturn size candidate planet in what was originally reported as a 66-day orbit. Subsequent observations have resulted in 3 additional observations of GJ3470-c, which led to an update of the orbital period to 32.998-days.

The original GJ3470-c submission to Arxiv.org included several additional uncharacterized transit observations. These included two of the transits used to characterize GJ3470-d. The Habitable Exoplanet Project ceased active observations in 2021, but Phillip Scott continued an active observing program of GJ3470 from that time to present. These observations resulted in five additional transits of GJ3470-d, for a total of seven, with excellent timing results over a 3-year period. In addition to GJ3470-d, we have observed five transits of a smaller planet, GJ3470-e, with very nearly the same orbital period.

# 2 Methodology

The method used to discover our pair of exoplanet candidates is transit photometry, which detects changes in brightness of a star potentially caused by an exoplanet transiting between the parent star and our position. Most of the data was collected using a 12.5" f5 Newtonian telescope at the OKSky Observatory. Exposures were 60 seconds with a luminance filter and an SBIG ST-8 camera. One detection was made using the Beatty 1-meter telescope at New Mexico Skies Observatory. Exposures were 40 seconds with luminance filter and an SBIG STXL-16803 camera. One detection was made using a C-14 f6.3 telescope at Jaxon Taylor Observatory in Oklahoma City. The images were 30 seconds with a ASI294mm Pro camera with luminance filter. That transit was also observed by Jim Edlin from Arizona. He was shooting 60 second images with a C-14 telescope and SBIG-ST8300 CCD camera.

Calibration frames were used to process images and then analyzed using AstroImageJ (AIJ) software. Comparison stars were selected from those recommended by the 'AAVSO Variable Star

Plotter' (AAVSO, 2023) for analysis, when possible, as long as no unusual variability was evident. Annulus settings recommended by AIJ were used and detrend for Airmass was applied. Analysis with AIJ was performed using published standards recommended by Dennis Conti (Conti, 2022) of AAVSO. AstroImageJ is a popular software developed by the University of Louisville for astronomical image analysis and precise photometry (University of Louisville, 2020). All of the observations were made between December, 2019 and April, 2023.

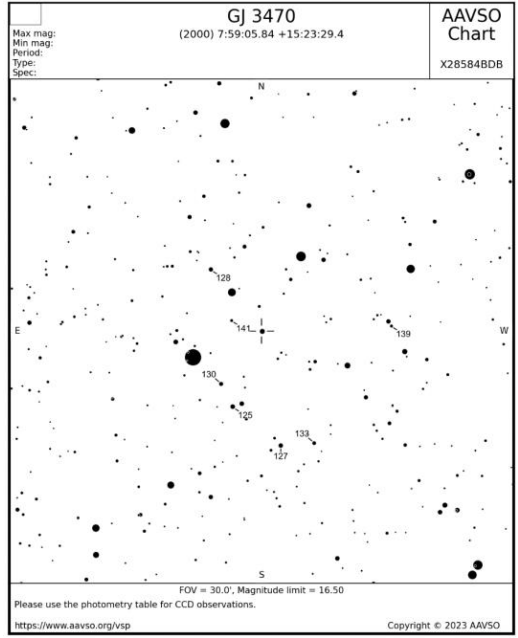

Figure 1 – Comparison stars for GJ3470 (AAVSO, 2023)

The data below corresponds to our target star GJ3470 (NASA, 2020).

| Distance | 30.7 Pc |
| --- | --- |
| Spectral Type | M1.5 |
| Apparent magnitude | 12.33 |
| Mass | 0.539 Solar masses |
| Radius | 0.547 Solar radii |
| Metallicity | 0.2 Fe/H |
| RA | 07:59:05.84 |
| Dec | +15:23:29.4 |

Table 1: GJ3470 Information

# 3 Transit photometry

## Candidate GJ3470-d

Six of the seven transits for candidate GJ3470-d were observed from the OKSky Observatory located in Oklahoma, USA (GPS coordinates -95:53:56, +34:41:24, 250 m). The Beatty 1-meter telescope at NM Skies Observatory, USA (GPS coordinates -105:32, +32:54, 2250 m) captured the 1/30/2023 transit. The full set of transits observed spanned 1,164 days, which allowed for a very precise orbital period calculation. They have an average depth of 1.4% and duration of 3 hours, 4 minutes. Orbital period, based on our observations, is 14.9617-days. Additionally, the last transit was observed by three separate observers in Oklahoma and Arizona with very good agreement in results.

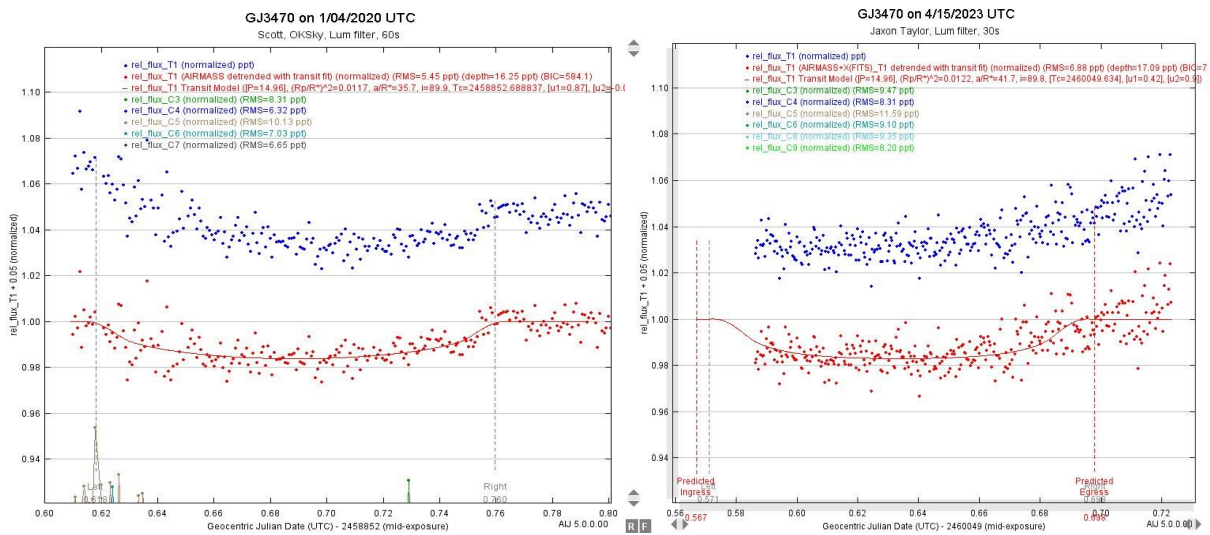

Figure 2: The first transit was detected by OKSky Observatory with a Tc of JD2458852.695. Duration was measured at 3h 25m at a depth of 1.6%. Source: OKSky Observatory
The partial transit on the right was detected by Jaxon Taylor Observatory with an estimated Tc of JD2460049.634. Estimated duration was 3h 3m at a depth of 1.7%. Source: Jaxon Taylor

| Calendar Date (UTC) | Julian Date | Tc (JD) | Transit Depth | Transit Duration (Hr/m) | Error (JD) | Observatory |
|---|---|---|---|---|---|---|
| 1/04/2020 | 2458852.695 | .695 | 1.6% | 3:25 | -0.004 | OKSky |
| 1/19/2020 | 2458867.661 | .661 | 1.8% | 3:13 | 0 | OKSky |
| 2/18/2020 | 2458897.571 | .571 | 1.3% | 2:59 | -0.004 | OKSky |
| 1/12/2021 | 2459226.776 | .779 | 1.8% | 3:14 | 0.035 | OKSky |
| 1/30/2023 | 2459974.824 | .824 | 0.8% | 2:56 | -.002 | NMskies |
| 3/01/2023 | 2460004.725 | .725 | 1.3% | 2:54 | -.022 | OKSky |
| 4/15/2023 | 2460049.641 | ≈.641 | 2.1% ? | 2:57? | 0.008? Partial | OKSky |
| 4/15/2023 | 2460049.634 | ≈.634 | 1.7% ? | 3:03? | 0.001? Partial | Jaxon Taylor |
| 4/15/2023 | 2460049.641 | ≈.641 | 0.9%? | ? | 0.008? Partial | Jim Edlin |

Table 2: GJ3470-d transits: Partial transit values with question marks were not used to calculate averages (4/15/2023 was three separate observations of the same transit)

# Candidate GJ3470-e

The five transits for candidate GJ3470-e were each observed from the OKSky Observatory located in Oklahoma, USA. The transits spanned 748 days. They have an average depth of 0.5% and duration of 3 hours, 2 minutes. Orbital period is 14.9467-days. Further attempts to observe GJ3470-e from Rio Cofio Observatory in Spain have been unsuccessful due to poor weather.

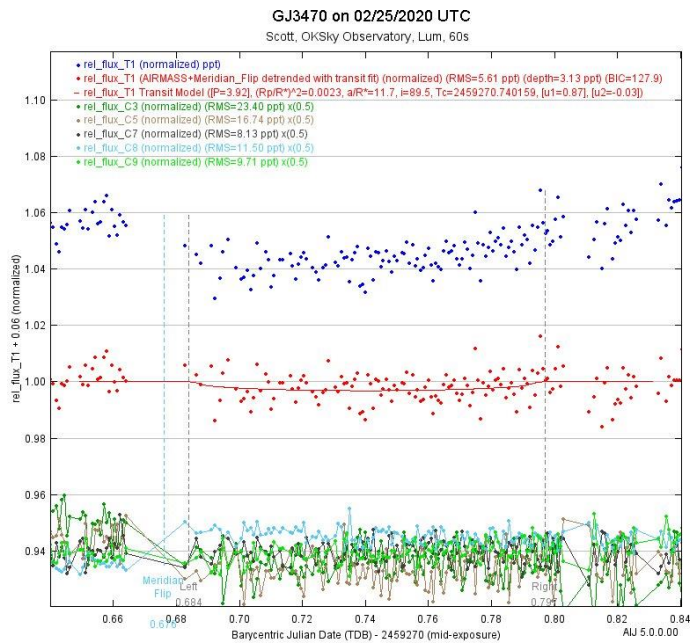

Figure 3: The first transit was observed on JD2459270.740.
Duration was measured to be 2h 51m at a depth of 0.35%. Source: OKSky Observatory

| Calendar Date (UTC) | Julian Date | Tc (JD) | Transit Depth | T. Duration (Hr/m) | Lead Time (JD) Relative to GJ3470-d |
|---|---|---|---|---|---|
| 2/25/2021 | 2459270.740 | .740 | 0.35% | 2:51 | 0.886 |
| 1/14/2023 | 2459958.771 | .771 | 0.8% | 3:19 | 1.092 |
| 2/13/2023 | 2459988.666 | .666 | 0.5% | 3:22 | 1.121 |
| 2/28/2023 | 2460003.661 | .661 | 0.4% | 2:44 | 1.137 |
| 3/15/2023 | 2460018.564 | .564 | 0.5% | 2:59? | 1.146 |

Table 3: GJ3470-e transits and lead time on GJ3470-d

# 4 Discussion

Our candidate, GJ3470-d, has an observed orbital period of 14.9617-days, but does not rule out shorter periods, if divided by whole numbers. However, since transit durations are already very near the maximum duration of 3 hours, 4.5 minutes for a 14.9617-days orbit, we believe the value of 0.09671 AU for the semi-major axis to be correct. There were considerable gaps in the data due to weather, technical issues, or other circumstances. Some possible transits were ruled out due to excessively short durations, incomplete (partial) transits, or unreasonable depths. Transits that were found acceptable have very small timing errors and have reasonable depths and durations.

Transits of the second candidate, GJ3470-e, have been observed over a 2-year span allowing us to calculate an orbital period of 14.9467-days. The similarity of the orbital periods of GJ3470-d and GJ3470-e (see figure 11 below) lead us to conclude that the two planets are co-orbital. GJ3470-e has been observed to lead GJ3470-d by about 1.1-days.

Our first consideration was that a smaller planet was caught in the LaGrange point, L4. For this to be the case, the orbital period would need to be half of the current observed period of 14.9467-days to meet the approximate 60-degree position requirement for L4. A 7.5-days orbit is doubtful due to the excessive transit durations. The maximum transit duration for a 7.5-days orbit was determined to be 2 hours, 24 minutes (JD 0.0102), considerably less than what has been observed. Also, a stable L4 Lagrange point requires a 1:25 mass ratio for the two objects. Ours is roughly estimated to be a 1:5 mass ratio, which is far too small for a stable position (Kemp, 2015).

Another possibility to consider is that GJ3470-e is an exomoon. However, we did not observe any significant changes in position relative to GJ3470-d. As a moon, we would expect to see it on both sides of the parent object, or at least significant changes in position. If the moon's orbit was in sync with the parent object, it might be possible to observe it in the same approximate position. However, it was always observed on the same side of its parent object at about four times the radius needed for an approximate 15-days orbit in sync with the parent's orbit around the star. We therefore conclude that GJ3470-e is unlikely to be a moon (Kipling, 2022).

Our next consideration was that of two planets in a horseshoe exchange orbit. An exchange orbit occurs when two co-orbital objects are of similar masses and thus exert a non-negligible influence on each other. They can swap eccentricities or semi-major axis (Funk, 2010). GJ3470-e appears to meet all of the requirements for a horseshoe exchange orbit. It is currently leading by 1.146-days and is gradually increasing its lead by roughly 7.5 minutes/orbit (JD 0.00521). An example of a Horseshoe Exchange orbit is found in our own solar system with two moons of Saturn. Epimetheus and Janos have been found orbiting Saturn in a horseshoe exchange of orbits about every 4-years (Funk, 2010). GJ3470-e currently has an estimated velocity that is 0.03 km/s faster than GJ3470-d. This means that GJ3470-e should catch up with GJ3470-d on the other side of the horseshoe in less than 8-years.

The table below shows a sine-wave graph that represents an assumed circular orbit of GJ3470-d. Shown are the plots of all five GJ3470-e transit detections and the last three GJ3470-d detections. The GJ3470-e's lead has increased from 0.886-day to 1.146-days (6 hour, 14.4 minutes) over the course of 50 orbital cycles for an average increase 7.49-minutes per orbit.

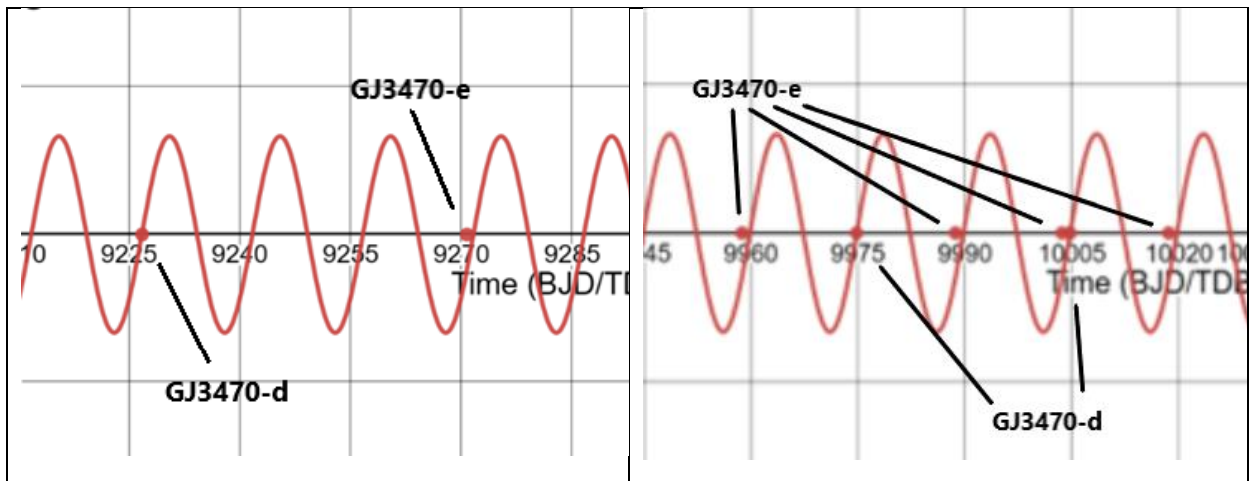

Figure 4: GJ3470-e was observed at JD2459270.740, JD2459958.771, JD2459988.666, JD2460003.611. It was always observed leading GJ3470-d. (Desmos online graphing calculator, 2023)

# 5 Characterization of candidates GJ3470-d and GJ3470-e

From the observed data we calculated the radius of each exoplanet candidate. The following equation gives a radius of **46,159 km** for GJ3470-d and **27,188 km** for GJ3470-e (Czernia, 2023).

$$D = (r/R)^2$$

$$r = R \cdot \sqrt{D}$$

*D = transit depth
*r = planet radius
*R = star radius

We calculated the semi-major axis of our candidates. Since the two planets share nearly the same orbit, the semi-major axis will very nearly be the same for both. Considering a stellar mass of 0.539 solar masses, a period of 14.9617-days, and neglecting planet mass we can obtain a close approximation of the semi-major axis for GJ3470-d of **0.09671 AU**. For GJ3470-e and its 14.9467-days orbit the semi-major axis is **0.09665 AU** (Czernia, 2023).

$$a \approx \sqrt[3]{GM_s(P/2\pi)^2}$$

*a = semi-major axis
*G = 6.67428 x 10⁻¹¹ m³ kg⁻¹ s⁻²
*Ms = 0.539 · 1.989 x 10³⁰ kg
*P = 14.9617-days

Orbital velocity was calculated with the equation below. The values assume a circular orbit and GJ3470-d has a value of **70.32 km/s**. The value calculated for GJ3470-e is **70.35 km/s**. (Czernia, 2023)

$$v = 2\pi a/P$$

v = velocity
a = Semi-major axis
P = Orbital period

Knowing the velocity allows us to use the following equation to calculate the maximum transit duration (Haswell, 2010).

$$T_{dur} = (P/\pi) \sin^{-1}(R_s/a)$$

T = Max duration
P = Orbital period
R = 380,898 km
a = semi-major axis

The maximum duration is **JD 0.1282** = 3 hours, 4.5 minutes. Our average duration of 3 hours, 4 minutes indicates that the inclination is very close to 90 degrees. In other words, GJ3470-d transits it's star almost dead center

The volume of each planet was calculated using the equation below (Haswell, 2010).

$$V = \frac{4}{3} \pi r^3$$

The volume of GJ3470-d was found to be **4.12 x 10$^{14}$ km³**. The volume of GJ3470-e was found to be **8.41 x 10$^{13}$ km³**

If we assume roughly identical densities, then we can assume a mass ratio of about **4.9:1**. This is far below the ratio currently accepted for stability of an object in a Lagrange point (Kemp, 2015).

# 6 Conclusion

We report the discovery of two new exoplanets orbiting GJ3470, an M1.5 red dwarf in Cancer. The exoplanets were observed co-orbiting in a Horseshoe Exchange orbit. GJ3470-d is the larger of the two with a measured diameter of 92,318 km. It has a transit duration of 3 hours, 4 minutes, a depth of 1.4%, and an orbital period of 14.9617-days. The semi-major axis is 0.09671 AU. Assuming a circular orbit, the orbital velocity is 70.32 km/s. The maximum transit duration for that orbital distance across GJ3470 was calculated as 3h 4.5m. Our average duration of 3h 4m, means inclination is very near 90 degrees.

The second exoplanet, GJ3470-e, is the smaller of the two. Its diameter was calculated to be 54,376 km and leads the larger planet by 1.146-days (1d 3h 30m) at the date of last observation. Data shows that GJ3470-e's lead distance is increasing by 7.49-minutes during each orbit of GJ3470-d. These will be the second and third exoplanets out of three exoplanets discovered and characterized by amateur astronomers without professional data or assistance. All three exoplanets, GJ3470-c, GJ3470-d, and GJ3470-e, were amateur discoveries made by Scott from his OKSky Observatory located in Oklahoma. When confirmed, this will likely be the first ever discovery of co-orbiting exoplanets and in particular, the first objects outside the solar system that co-orbit in a Horseshoe Exchange orbit.